\documentclass[10pt,notitlepage,showpacs,pra,twocolumn]{revtex4-1}
\usepackage{ulem}
\usepackage{amsfonts}
\usepackage{amsmath}
\usepackage{amssymb}
\usepackage{graphicx}
\usepackage{float}
\usepackage{chemarrow}
\usepackage[toc,page,header]{appendix}

\begin{document}

\title{Spheroidal-structure-based multi-qubit Toffoli gate via asymmetric Rydberg interaction}
\author{Dongmin Yu$^{1}$, Weiping Zhang$^{2,3}$, Jin-ming Liu$^{1}$, Shilei Su$^{4,*}$ and Jing Qian$^{1,\dagger}$ }
\affiliation{$^{1}$State Key Laboratory of Precision Spectroscopy, Department of Physics, School of Physics and Electronic Science, East China
Normal University, Shanghai 200062, China}
\affiliation{$^{2}$Department of Physics and Astronomy, Shanghai Jiaotong University and Tsung-Dao Lee Institute, Shanghai 200240, China}
\affiliation{$^{3}$Collaborative Innovation Center of Extreme Optics, Shanxi University, Taiyuan, Shanxi 030006, China}
\affiliation{$^{4}$School of Physics, Zhengzhou University, Zhengzhou 450001, China}

\begin{abstract}
We propose an exotic multi-qubit Toffoli gate protocol via asymmetric Rydberg blockade, benefiting from the use of a spheroidal configuration to optimize the gate performance. The merit of a spheroidal structure lies in a well preservation of strong blocked energies between all control-target atom pairs within the sphere, which can persistently keep the blockade error at a low level. On the basis of optimization for three different types of $(2+1)$-$qubit$ gate units to minimize the antiblockade error, the gate fidelity of an optimal $(6+1)$-$qubit$ configuration can attain as high as $0.9841$ mainly contributed by the decay error. And the extension with much more control atoms is also discussed. Our findings may shed light on scalable neutral-atom quantum computation in special high-dimensional arrays.

\end{abstract}
\email{jqian1982@gmail.com}
\pacs{}
\maketitle
\preprint{}

\section{Introduction}

Multi-qubit quantum gate lies in the heart of universal quantum computing and is able to speedup the quantum algorithms and state preparation in complex systems\cite{nielsen_2010}. While the fact that implementation of a single three-qubit Toffoli gate requiring sequence multiple control-NOT gates makes the circuit system complexity \cite{Barenco_1995,Ralph_2007,Yu_2013,Biswal_2019}. To make matters worse, the required number of decomposed two-qubit gate rises exponentially as the qubit number of multi-qubit gate increases. Thus, to construct the multi-qubit quantum gate directly has practical significance in reducing the complexity of quantum circuits. Fortunately, the long-range nature of Rydberg-atom blockade interactions has the potential to manipulate single target atom by multiple control atoms, applying for an achievement of multi-qubit Toffoli gate \cite{Saffman_2016,Shen_19}. However this protocol still stays at the theoretical level due to the relatively low quantum state initialization and ground-state coherent control. Three-body F\"{o}rster resonance was applied to facilitate the interatomic dipole-dipole interactions allowing for a fast and high-fidelity three-qubit Toffoli gate \cite{Beterov_2018}. Until very recently the Toffoli gate based on Rydberg blockade was first demonstrated experimentally with atoms trapped in a 1D array of optical tweezers, leading to a fidelity of $\geq0.837$ \cite{Levine_2019}. But such a 1D scheme is uneasy for extending into a scalable multi-qubit quantum computing network because the long-range next-to-nearest control-target interaction is too weak to engineer the target qubit. 

In the present work we show the universal production of a $(6+1)-qubit$ Toffoli gate via asymmetric Rydberg blockade (ARB), in which the control-target dipole-dipole(DD) interaction is much larger than the control-control van der Waals(vdWs) interaction \cite{Saffman_2009,Su_2018,2020arXiv200602486Y,Wu_2010}. It is realized by exciting control or target atoms into different Rydberg levels and has been applied for {\it e.g.} state control \cite{Brion_2007} or entangled state preparation \cite{Carr_2013}.
However, unlike previous schemes that depend on 1D linear or 2D square arrays,  
our most efficient protocol adopts a blockade sphere configuration with all control atoms positioned on the surface and one target atom in the spherical center. This novel configuration can ensure a preserved blockaded energy between the control-target atoms accomplished by a proper spherical radius, which is comparable to the Rydberg blockade radius. On that basis, we only need to consciously care about how to modulate the control-control interactions. Accompanied by an optimization of the control-atom position in basic $(2+1)$-$qubit$ gate units, a large difference of scales between the DD and vdWs Rydberg interactions can be realized, deeply reducing the anitiblockade error from a non-zero vdWs interaction. Finally we study the robustness of a spheroidal $(6+1)$-$qubit$ gate by adding the 7th control atom with a stochastic position on the surface, and reveal its sustained high fidelity as long as extra control atoms are suitably placed.

\section{Asymmetric Rydberg blockade and our spheroidal gate} 
The concept of asymmetric blockaded energy in production of multi-qubit Rydberg gates has been widely utilized. Because in the situation of ARB, where the control-control interaction $U_{cc}$ is much smaller than the exciting laser Rabi frequency $\Omega$, all control atoms can be excited simultaneously; while the sufficient control-target interaction $U_{ct}$ is able to block the target atom's excitation, arising the production of a so-called $C_kNOT$ gate(also known as multi-qubit Toffoli gates). However such a large asymmetry of $U_{ct}\gg\Omega\gg U_{cc}$ is hard to meet in conventional 1D or 2D Rydberg atom chains. 
Because it is unable to preserve a steady strong blockade interaction $U_{ct}$ as $k$(the number of control atoms) grows leading to a weaker control-target engineering. To solve this difficulty, originating from the idea of blockade sphere(``superatom") \cite{PhysRevX.5.031015,np.11.157} we propose a spheroidal-structured multi-qubit gate which benefits from a preserved strong blocked energy $U_{tc}$ for all control-target interactions. Such a structure can steadily maintain the blockade error at a lower level. Thus, one can safely optimize the control-control spacing to minimize $U_{cc}$, which would deeply reduce the antiblockade error and further achieve the higher gate fidelity. 

\begin{figure}[htbp]
\centering
\fbox{\includegraphics[width=\linewidth]{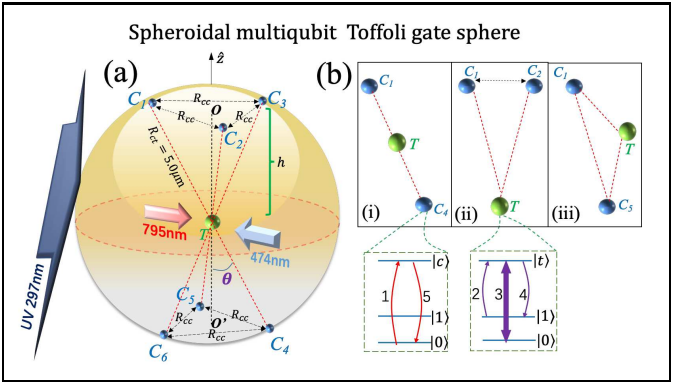}}
\caption{(a) Global schematic of a spheroidal $(6+1)$-$qubit$ Toffoli gate, 
including one target atom and six control atoms. To be specific three control atoms $C_{1,2,3}$($C_{4,5,6}$) structuring of an equilateral are placed on the spherical surface with the unique target atom $T$ in the center of sphere. $h$ stands for the vertical height between the control atom and the plane formed by control atoms. $\hat{z}$ is set to be a quantization axis controlled by the external field that induces a DD control-target interaction. (b) Three basic types of $(2+1)$-$qubit$ gate involved in this multi-qubit spheroidal gate, which are (i) linear type, (ii) acute triangle, (iii) obtuse triangle. 
Insets: Level scheme of control and target atoms interacting with pulses $1\sim5$. }
\label{mod}
\end{figure}

With this expectation we consider the situation shown in Fig. \ref{mod} where a $(6+1)$ atom qubit with states $|0\rangle,|1\rangle$ encoded in hyperfine ground substates and $|c\rangle$($|t\rangle$) for Rydberg states of control(target) atoms. The conventional implementation of basic $(2+1)$-$qubit$(Toffoli) gate units like (i)-(iii) in Fig. \ref{mod}(b) requires three fundamental steps \cite{Shi_2018}: (1) The incidence $\pi$ pulse with Rabi frequency $\Omega_1$ is applied to control atoms, allowing a full Rydberg excitation $|0\rangle\to|c\rangle$. (2) A pair of $\pi$ pulses $\Omega_{2\sim4}$ that implement on the target atom, can cause a reversible state exchange between $|0\rangle\rightleftarrows|t\rangle\rightleftarrows|1\rangle$ if all control atoms are idler in $|1\rangle$. Here the direction of population transfer caused by $\Omega_{3}$ depends on the initial state is $|0\rangle$ or $|1\rangle$. (3) A $(-\pi)$-pulse $\Omega_5$ returns the control atoms back into $|0\rangle$ from $|c\rangle$ via a de-excitation process. Note that $\Omega_{c(t)}$ are the amplitudes of control (target) laser Rabi frequencies.
Ideally when the incident qubits are $|1...1\beta\rangle$ meaning all control atoms are idler, the target state $|\beta\rangle$ will pick up a state exchange $|\beta\rangle\to|\bar{\beta}\rangle$ with $\beta\in[0,1]$ and $\bar{\beta}\equiv1-\beta$. Otherwise all states remain unchanged. So the desired gate transformation will follow the route of: $|00\beta\rangle\xrightarrow{\Omega_1}i|cc\beta\rangle\xrightarrow{\Omega_{2\sim4}}i|cc\beta\rangle\xrightarrow{\Omega_5}|00\beta\rangle$, $|10\beta\rangle\xrightarrow{\Omega_1}i|1c\beta\rangle\xrightarrow{\Omega_{2\sim4}}i|1c\beta\rangle\xrightarrow{\Omega_5}|10\beta\rangle$, $|01\beta\rangle\xrightarrow{\Omega_1}i|c1\beta\rangle\xrightarrow{\Omega_{2\sim4}}i|c1\beta\rangle\xrightarrow{\Omega_5}|01\beta\rangle$, $|11\beta\rangle\xrightarrow{\Omega_1}|11\beta\rangle\xrightarrow{\Omega_{2\sim4}}|11\bar{\beta}\rangle\xrightarrow{\Omega_5}|11\bar{\beta}\rangle$.

For realizing the desired multi-qubit Rydberg gate an asymmetric state-dependent blockade relies on the chosen of specific states based on $^{87}Rb$ atoms \cite{Isenhower_2011}: $|0\rangle=|5S_{1/2},F=1,m_F=0\rangle$, $|1\rangle=|5S_{1/2},F=2,m_F=0\rangle$, specifically Rydberg states are $|c\rangle=|61P_{3/2}\rangle$ and $|t\rangle=|61S_{1/2}\rangle$, which supports a strong dipole-dipole interaction energy $U_{ct}$. This blocked energy $U_{ct}=C_3^{ct}(\theta)/R_{ct}^3$ depends on the spatial orientation of control-target distance $R_{ct}$ where $C_3^{ct}(\theta)=C_3^{ct}(1-3\cos^2\theta)$ \cite{Petrosyan_2014}, the coefficient $C_3^{ct}/2\pi=4.73$~GHz$\mu$m$^3$ and $\theta$ is the angle between $\hat{z}$ and $\Vec{R}_{ct}$ ($R_{ct}=|\Vec{R}_{ct}|$). Given $R_{ct}=5.0~\mu$m, the adjustable blocked energy $U_{ct}(\theta)$ varies between $2\pi\times(-75.68,37.84)$~MHz. However the intraspecies control-control interaction between states $|c\rangle$ is ensured to be very weak by state property and of an isotropic vdWs-type, denoted by $U_{cc}=C_6^{cc}/R_{cc}^6$, satisfying an asymmetric nature $U_{cc}\ll U_{ct}$. Here $C_6^{cc}/2\pi =14.9 $~GHz$\mu$m$^6$ and $R_{cc}\in(0,2R_{ct})$ is adjustable via modulaitng \emph{h}. 
Moreover we find the control-control distance $R_{cc}$ leads to the energy $U_{cc}$ conflicted as $h$ varies between cases of (ii) and (iii). So it is necessary to search for a best $h$ value, enabling a maximal-fidelity implementation of different $(2+1)$-$qubit$ gate units at the same time. Note that $U_{ct}(\theta)$ also depends on $h$ because $\cos\theta=h/R_{ct}$.

\section{Optimal asymmetric blockade}
To achieve the desired asymmetric interactions and couplings we have considered the large separation of scales between different types of Rydberg interactions $U_{ct}$, $U_{cc}$ which both can be tuned by $h$. The effective Hamiltonian for describing the dissipative dynamics of a $(2+1)$-$qubit$ gate unit, is expressed as $\hat{\mathcal{H}}_{\rm eff}=\hat{\mathcal{H}}_0+\hat{\mathcal{H}}_I-\frac{i}{2}\sum_{k=c_p,c_q,t}\hat{\mathcal{L}}_k^{\dagger}\hat{\mathcal{L}}_k $,
 where $\hat{\mathcal{H}}_0= \hat{\mathcal{H}}_{c_p}+\hat{\mathcal{H}}_{c_q}+\hat{\mathcal{H}}_t=\frac{1}{2}\sum_{j=c_p,c_q}(\Omega_{1}\hat{\sigma}_{0c_j}+\Omega_{5}\hat{\sigma}_{c_j0})+\frac{1}{2}[\Omega_2\hat{\sigma}_{1t}+\Omega_4\hat{\sigma}_{t1}+\Omega_3(\hat{\sigma}_{0t}+\hat{\sigma}_{t0})]$ represents the time-dependent atom-light coupling and $\hat{\mathcal{H}}_I= U_{cc}\hat{\sigma}_{c_pc_p}\hat{\sigma}_{c_qc_q}+U_{ct}(\hat{\sigma}_{c_pc_p}\hat{\sigma}_{tt}+\hat{\sigma}_{c_qc_q}\hat{\sigma}_{tt})$ is the atom-atom Rydberg interaction. Here $\hat{\sigma}_{mn}=|m\rangle\langle n|$ and the subscripts $p,q\in(1,2,3,4,5,6)$, and $p\not=q$. We solve the motional dynamics of arbitrary three-qubit input state $|\Psi\rangle$[=\{$|000\rangle,|001\rangle,|100\rangle,|010\rangle,|101\rangle,|011\rangle,|110\rangle,|111\rangle$\}] by the Schr\"{o}dinger equation ($\hbar=1$): $ \partial_t|\Psi\rangle=-i\hat{\mathcal{H}}_{eff}|\Psi\rangle$
using the quantum stochastic wavefunction \cite{Molmer_1993}. After averaging over 500 evolutions of the Schr\"{o}dinger equation via the Monte Carlo simulation one obtains the final results. $\hat{\mathcal{L}}_{k}$ indicates the spontaneous dissipation of Rydberg levels, taking forms of $\hat{\mathcal{L}}_{k}=\sqrt{\Gamma_k}(|1\rangle\langle k|+|0\rangle\langle k|)$ where the decay rates are $\Gamma_k=\Gamma_c$ for $k=c_{p,q}$ and $\Gamma_k=\Gamma_t$ for $k=t$. In practice the control atoms are globally driven via a one-step UV excitation with wavelength 297nm; and the target atom will face a two-photon transition with wavelengths 795nm and 474nm, limited by the selection rules. For the target atom the intermediate state {\it e.g.} $|5P_{1/2}\rangle$ has been safely discarded due to dispersive interactions.

\begin{figure}[htbp]
\centering
\fbox{\includegraphics[width=\linewidth]{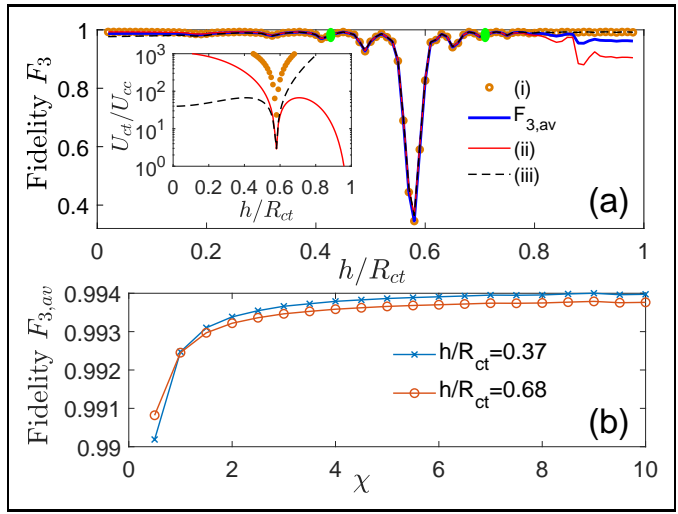}}
\caption{ (a) Fidelities $\mathcal{F}_3$ as a function of the height $h$ with respect to the blockade radius $R_{ct}=5~\mu$m, for different case (i) (orange dots), case (ii) (red-solid) and case (iii) (black-dashed). The average value $\mathcal{F}_{3,av}$ including the contributions from three cases is labeled by the thick blue-soild curve. The maximum of average value $\mathcal{F}_{3,av}=0.9925$ is highlighted by the green dots at $h/R_{ct}= 0.37$ and $0.68$, respectively. (b) Average fidelity $\mathcal{F}_{3,av}$ over three configurations (i)-(iii) using the parameters of (a), and $h/R_{ct}=0.37$(blue curve with crosses) and $0.68$ (red curve with circles).}
\label{fid}
\end{figure}

To develop an optimal protocol for a best ARB condition we proceed by numerically estimating the fidelity that depends on the strength of blockade energy. The fidelity averaging over all input states is given by $\mathcal{F}_{n}=\frac{1}{2^n}\text{Tr}[\sqrt{\rho_{et}}|\Psi\rangle\langle\Psi|(t=t_{det})\sqrt{\rho_{et}}]^{1/2}$, with $n$ the number of qubits (here $n=3$ means 2 control qubits and 1 target qubit) and $\rho_{et}$ an etalon matrix of ideal Toffoli gate. $t_{det}=2(\pi/\Omega_c)+3(\pi/\Omega_t)$ describes the detection time after all pulses. For simplicity $\Omega_0=\Omega_{c,t}=2\pi\times3.784$~MHz is used giving to $t_{det}=0.66~\mu$s. The dissipative rates are estimated by $\Gamma_c/2\pi=2.0$~kHz, $\Gamma_t/2\pi=4.0$~kHz by ref.\cite{Beterov_2009}. As shown in Fig. \ref{fid}a, we find that high fidelities of $F_3>0.95$ is kept for most height values. Except if $\theta=\cos^{-1}(1/\sqrt{3})$({\it i.e.} $h/R_{ct}=1/\sqrt{3}$) where the control-target interaction $U_{ct}$ has a zero that is unable to engineer the target atom, it arises a lowest deep below 0.4 of fidelity. The interaction asymmetry of three cases is compared in the inset of Fig. \ref{fid}a. It shows that the behavior of $U_{ct}/U_{cc}$ is quite opposite between cases (ii) and (iii) due to the structure dependence. 
However $U_{ct}/U_{cc}\gg1$ is mostly satisfied confirming that a large asymmetric interaction is always preserved in our model. Presently we will choose especially $h/R_{ct}=0.37$ or $0.68$ which ensures a maximal average fidelity $F_{3,av}=0.9925$ as denoted by the green dots in Fig. \ref{fid}a.

In addition, we find that the performance of the gate can be further improved by utilizing imbalanced coupling strengths {\it i.e.} $\chi=\Omega_c/\Omega_t\neq 1$. When $\Omega_t/2\pi=3.784$~MHz we vary $\Omega_c$ for a best average fidelity. For a larger $\Omega_c$ the antiblockade excitation rate between two control atoms is expected to be enhanced allowing for an improving fidelity due to the requirement of $\Omega_c\gg U_{cc}$ in antiblockade. As we illustrate in Fig. \ref{fid}b using the optimal heights $h/R_{ct}=0.37,0.68$ the average fidelity increases and tends to be saturation for a larger $\chi$. Since an asymmetric coupling strengths lead to an auxiliary enhancement to $\mathcal{F}_{3,av}$, in what follows, we show the implementation of a multi-qubit Toffoli gate with best parameters $h/R_{ct}=0.37,\Omega_c/\Omega_t=10$, and a more generalized condition for realizing ARB should satisfy  $U_{ct},\Omega_c\gg\Omega_t\gg U_{cc}$.

\section{Gate-error sources}

\begin{figure}[htbp]
\centering
\fbox{\includegraphics[width=\linewidth]{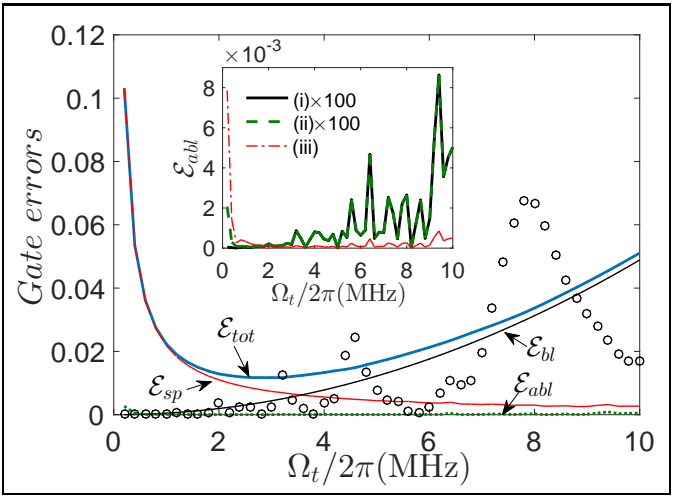}}
\caption{Error sources of the $(2+1)$-$qubit$ gate units {\it versus} Rabi frequency $\Omega_t$. The black dotted points are obtained from numerical simulations and the curve $\mathcal{E}_{bl}$(black-solid) is fitted by $\mathcal{E}_{bl}\approx0.7(\Omega_t/U_{ct})^2$. Inset represents a zoom-in image for the antiblockade error $\mathcal{E}_{abl}$ in the cases of (i)-(iii).}
\label{Fig4}
\end{figure}

To explore the error source of our scheme it is instructive to recall how the errors play roles for a symmetric Rydberg blockade gate. Relevant results are plotted in Fig. \ref{Fig4}. It can be seen that for a small Rabi frequency the dominant error is contributed by the spontaneous decay of Rydberg states which can be  analytically expressed as $\mathcal{E}_{sp}\approx \pi\Gamma_t/4\Omega_t+3\pi\Gamma_c/\Omega_t+\pi\Gamma_c/\Omega_c$. In our scheme by substituting parameters we find $\mathcal{E}_{sp}\approx 0.006$ perfectly agreeing with the best average fidelity obtained in Fig. \ref{fid}b. As increasing $\Omega_t$ the blockade error $\mathcal{E}_{bl}$ quadratically grows, approximately fitted by $\mathcal{E}_{bl}\approx0.7(\Omega_t/U_{ct}(\theta))^2$(for $h/R_{ct}=0.37$, $\theta=68.28^{o}$, black-solid curve) originating from the insufficient control-target blockaded interaction. We emphasize that the original blockade error shown in Fig. \ref{Fig4} (black-dots) is modified to be oscillating with Rabi frequency $\Omega_t$ \cite{PhysRevA.85.042310}. So the blockade error $\mathcal{E}_{bl}$ can be minimized by choosing an optimum value $\Omega_t$ letting $\mathcal{E}_{bl}\ll\mathcal{E}_{sp}$.

Differing from a conventional blockade gate, our scheme is also affected by the antiblockade error coming from the imperfect requirement of $\Omega_c\gg U_{cc}$. If the intraspecies interaction $U_{cc}$ is too big to permit the simultaneous excitation of control atoms it will lower the fidelity. Luckily since $\mathcal{E}_{abl}\approx(U_{cc}/\Omega_c)^2$ is two orders of magnitude smaller than $(U_{cc}/\Omega_t)^2$ , the influence of antiblockade error $\mathcal{E}_{abl}$ is also negligible here. A zoom-in image for detailed antiblockade errors is presented in the inset of Fig. \ref{Fig4}. In general $\mathcal{E}_{abl}$ can be made orders of magnitude smaller than other errors, benefiting from an optimal design of the spherical configuration. Therefore the total error $\mathcal{E}_{tot}\approx\mathcal{E}_{sp}+\mathcal{E}_{bl}$ (blue-solid) is mainly contributed by the two former errors. Since $\mathcal{E}_{sp}$ and $\mathcal{E}_{bl}$ do not depend on structures, the current scheme provides us with sufficient room to study the production of a muti-qubit Toffoli gate
immuned to the influence from imperfect asymmetric blockade.

\section{Spheroidal gate and its extension}

A generalized form of the decay error is given by $\mathcal{E}_{k,sp}\approx \frac{1}{2^k} (\pi\Gamma_t/\Omega_t)+k (3\pi\Gamma_c/2\Omega_t)+k(\pi\Gamma_c/2\Omega_c)$ \cite{Isenhower_2011} that increases with $k$. For $k=6$ it leads to $\mathcal{E}_{6,sp}\approx 0.0155$. By numerically solving the gate fidelity $\mathcal{F}_{7}$ following the Schr\"{o}dinger equation we obtain a best $(6+1)$-$qubit$ spheroidal gate fidelity $\mathcal{F}_{7}\approx0.9841$. This value is close to $1-\mathcal{E}_{6,sp}=0.9845$, indicating that the dominant error of scheme comes from the Rydberg-state decays. The other two errors have been deeply minimized to a very low level under optimization.

\begin{figure}[htbp]
\centering
\fbox{\includegraphics[width=\linewidth]{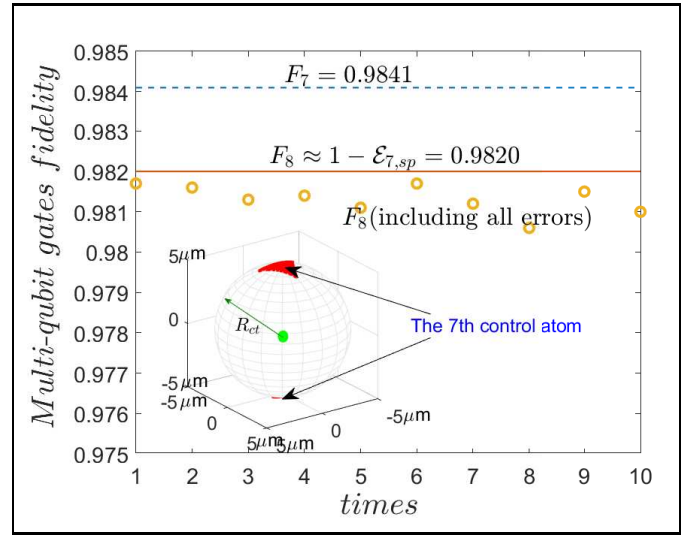}}
\caption{Multi-qubit spheroidal gate fidelity. The best $(6+1)$-$qubit$ gate fidelity is $F_7=0.9841$(blue-dashed) under optimization. The $(7+1)$-$qubit$ gate fidelity denoted by $F_8$, solved from the analytical expression $1-\mathcal{E}_{7,sp}$ (red-solid) and numerically obtained from the Schr\"{o}dinger equation (orange dots) with a stochastic position of the $7th$ control atom. Inset: schematic diagram of a spheroidal quantum gate in which the $7th$ control atom is stochastically placed at the pole areas of the sphere.}
\label{Fig5}
\end{figure}

Furthermore the robustness of such a spheroidal multi-qubit gate can be verified as randomly placing the $7th$ control atom on the spherical surface. Intuitively if the intraspecies interaction $U_{cc}$ is enhanced due to the position of the $7th$ atom closing to other atoms, the antiblockade error $\mathcal{E}_{abl}$ may become dominant, breaking the gate performance. To avoid this we seek for suitable positions on the surface that can keep the low level of antiblockade interaction. Here we assume $U_{cc}/2\pi<1$~MHz is required. With this constraint the $7th$ control atom is expected to be randomly positioned just on the pole areas of the sphere. For comparison, in Fig. \ref{Fig5} we plot the multi-qubit gate fidelity under different cases. Without extra atoms this $(6+1)$-$qubit$ gate permits a high-fidelity $F_7=0.9841$(blue-dashed). Intuitive speaking the additional $7th$ control atom would lower the gate fidelity to an ideal value $F_8=0.9820$(red-solid) accounting for an inevitable decay error that increases as $k$. However a practical multi-qubit gate implementation suffering from blockade and antiblockade errors, will reveal a lower value. By randomly positioning the $7th$ control atom for 10 times we numerically obtain the practical gate-fidelity $F_8$ (including all errors) with an average value of $\sim 0.9813$. This slight difference $0.0007$ from the ideal value, is determined by imperfect errors $\mathcal{E}_{bl}$ and $\mathcal{E}_{abl}$. Remarkably $\mathcal{E}_{bl}+\mathcal{E}_{abl}\ll\mathcal{E}_{k,sp}$($\mathcal{E}_{k,sp}\sim0.018$ for $k=7$) confirms the robustness of our spheroidal gate protocol that makes all errors except the decay error less important.

\section{Conclusion}

We have presented a spheroidal multi-qubit Toffoli gate scheme involving 1 target atom and 6 control atoms, enabled by the ARB mechanism. This newly-proposed spheroidal-structure-based ARB, to our knowledge, is found to have many features. First it allow for a perfect preservation of strong blockaded energy between any control-target atom pairs, sustaining a low-level blockade error which is almost unaffected by the number of control atoms. 
Second, under the optimization for the spacial positions of control atoms the antiblockade error of inter-control-atoms could be greatly reduced, with its value orders of magnitude smaller than other errors. Therefore this $(6+1)$-$qubit$ gate protocol is easier to be extended into a case with more control atoms as long as the interatomic distance is suitably chosen. Finally, after balancing of the requirement of strong blocked interaction and weak anti-blockded interaction we have realized a promising spheroidal multi-qubit gate scheme with a fidelity of $\sim0.98$. Straightforward extensions of this scheme can be used for the production of Rydberg-mediated multi-particle entanglement \cite{Saffman_2009,Graham_2019} or mesoscopic atomic ensembles \cite{PhysRevLett.102.170502}, and for fast quantum computation with neutral Rydberg qubits \cite{PhysRevLett.85.2208}.

\bigskip

\acknowledgements

This work was supported by the National Natural Science Foundation of China under Grants Nos. 11474094, 11104076, 11804308, 91950112, 11174081; the China Postdoctoral Science Foundation Grant No. 2018T110735, the Science and Technology Commission of Shanghai Municipality under Grant No. 18ZR1412800, the National Key Research and Development Program of China under Grant No. 2016YFA0302001, and the ECNU Academic Innovation Promotion Program for Excellent Doctoral Students under Grant No. YBNLTS2019-023.

\appendix

\bigskip

\bibliography{references}

\begin{thebibliography}{26}%
\makeatletter
\providecommand \@ifxundefined [1]{%
 \@ifx{#1\undefined}
}%
\providecommand \@ifnum [1]{%
 \ifnum #1\expandafter \@firstoftwo
 \else \expandafter \@secondoftwo
 \fi
}%
\providecommand \@ifx [1]{%
 \ifx #1\expandafter \@firstoftwo
 \else \expandafter \@secondoftwo
 \fi
}%
\providecommand \natexlab [1]{#1}%
\providecommand \enquote  [1]{``#1''}%
\providecommand \bibnamefont  [1]{#1}%
\providecommand \bibfnamefont [1]{#1}%
\providecommand \citenamefont [1]{#1}%
\providecommand \href@noop [0]{\@secondoftwo}%
\providecommand \href [0]{\begingroup \@sanitize@url \@href}%
\providecommand \@href[1]{\@@startlink{#1}\@@href}%
\providecommand \@@href[1]{\endgroup#1\@@endlink}%
\providecommand \@sanitize@url [0]{\catcode `\\12\catcode `\$12\catcode
  `\&12\catcode `\#12\catcode `\^12\catcode `\_12\catcode `\%12\relax}%
\providecommand \@@startlink[1]{}%
\providecommand \@@endlink[0]{}%
\providecommand \url  [0]{\begingroup\@sanitize@url \@url }%
\providecommand \@url [1]{\endgroup\@href {#1}{\urlprefix }}%
\providecommand \urlprefix  [0]{URL }%
\providecommand \Eprint [0]{\href }%
\providecommand \doibase [0]{http://dx.doi.org/}%
\providecommand \selectlanguage [0]{\@gobble}%
\providecommand \bibinfo  [0]{\@secondoftwo}%
\providecommand \bibfield  [0]{\@secondoftwo}%
\providecommand \translation [1]{[#1]}%
\providecommand \BibitemOpen [0]{}%
\providecommand \bibitemStop [0]{}%
\providecommand \bibitemNoStop [0]{.\EOS\space}%
\providecommand \EOS [0]{\spacefactor3000\relax}%
\providecommand \BibitemShut  [1]{\csname bibitem#1\endcsname}%
\let\auto@bib@innerbib\@empty
\bibitem [{\citenamefont {Nielsen}\ and\ \citenamefont
  {Chuang}(2010)}]{nielsen_2010}%
  \BibitemOpen
  \bibfield  {author} {\bibinfo {author} {\bibfnamefont {M.~A.}\ \bibnamefont
  {Nielsen}}\ and\ \bibinfo {author} {\bibfnamefont {I.~L.}\ \bibnamefont
  {Chuang}},\ }\href {\doibase 10.1017/CBO9780511976667} {\emph {\bibinfo
  {title} {Quantum Computation and Quantum Information}}}\ (\bibinfo
  {publisher} {Cambridge University Press},\ \bibinfo {year}
  {2010})\BibitemShut {NoStop}%
\bibitem [{\citenamefont {Barenco}\ \emph {et~al.}(1995)\citenamefont
  {Barenco}, \citenamefont {Bennett}, \citenamefont {Cleve}, \citenamefont
  {DiVincenzo}, \citenamefont {Margolus}, \citenamefont {Shor}, \citenamefont
  {Sleator}, \citenamefont {Smolin},\ and\ \citenamefont
  {Weinfurter}}]{Barenco_1995}%
  \BibitemOpen
  \bibfield  {author} {\bibinfo {author} {\bibfnamefont {A.}~\bibnamefont
  {Barenco}}, \bibinfo {author} {\bibfnamefont {C.~H.}\ \bibnamefont
  {Bennett}}, \bibinfo {author} {\bibfnamefont {R.}~\bibnamefont {Cleve}},
  \bibinfo {author} {\bibfnamefont {D.~P.}\ \bibnamefont {DiVincenzo}},
  \bibinfo {author} {\bibfnamefont {N.}~\bibnamefont {Margolus}}, \bibinfo
  {author} {\bibfnamefont {P.}~\bibnamefont {Shor}}, \bibinfo {author}
  {\bibfnamefont {T.}~\bibnamefont {Sleator}}, \bibinfo {author} {\bibfnamefont
  {J.~A.}\ \bibnamefont {Smolin}}, \ and\ \bibinfo {author} {\bibfnamefont
  {H.}~\bibnamefont {Weinfurter}},\ }\href {\doibase 10.1103/PhysRevA.52.3457}
  {\bibfield  {journal} {\bibinfo  {journal} {Phys. Rev. A}\ }\textbf {\bibinfo
  {volume} {52}},\ \bibinfo {pages} {3457} (\bibinfo {year}
  {1995})}\BibitemShut {NoStop}%
\bibitem [{\citenamefont {Ralph}\ \emph {et~al.}(2007)\citenamefont {Ralph},
  \citenamefont {Resch},\ and\ \citenamefont {Gilchrist}}]{Ralph_2007}%
  \BibitemOpen
  \bibfield  {author} {\bibinfo {author} {\bibfnamefont {T.~C.}\ \bibnamefont
  {Ralph}}, \bibinfo {author} {\bibfnamefont {K.~J.}\ \bibnamefont {Resch}}, \
  and\ \bibinfo {author} {\bibfnamefont {A.}~\bibnamefont {Gilchrist}},\ }\href
  {\doibase 10.1103/PhysRevA.75.022313} {\bibfield  {journal} {\bibinfo
  {journal} {Phys. Rev. A}\ }\textbf {\bibinfo {volume} {75}},\ \bibinfo
  {pages} {022313} (\bibinfo {year} {2007})}\BibitemShut {NoStop}%
\bibitem [{\citenamefont {Yu}\ \emph {et~al.}(2013)\citenamefont {Yu},
  \citenamefont {Duan},\ and\ \citenamefont {Ying}}]{Yu_2013}%
  \BibitemOpen
  \bibfield  {author} {\bibinfo {author} {\bibfnamefont {N.}~\bibnamefont
  {Yu}}, \bibinfo {author} {\bibfnamefont {R.}~\bibnamefont {Duan}}, \ and\
  \bibinfo {author} {\bibfnamefont {M.}~\bibnamefont {Ying}},\ }\href {\doibase
  10.1103/PhysRevA.88.010304} {\bibfield  {journal} {\bibinfo  {journal} {Phys.
  Rev. A}\ }\textbf {\bibinfo {volume} {88}},\ \bibinfo {pages} {010304}
  (\bibinfo {year} {2013})}\BibitemShut {NoStop}%
\bibitem [{\citenamefont {Biswal}\ \emph {et~al.}(2019)\citenamefont {Biswal},
  \citenamefont {Bhattacharjee}, \citenamefont {Chattopadhyay},\ and\
  \citenamefont {Rahaman}}]{Biswal_2019}%
  \BibitemOpen
  \bibfield  {author} {\bibinfo {author} {\bibfnamefont {L.}~\bibnamefont
  {Biswal}}, \bibinfo {author} {\bibfnamefont {D.}~\bibnamefont
  {Bhattacharjee}}, \bibinfo {author} {\bibfnamefont {A.}~\bibnamefont
  {Chattopadhyay}}, \ and\ \bibinfo {author} {\bibfnamefont {H.}~\bibnamefont
  {Rahaman}},\ }\href {\doibase 10.1103/PhysRevA.100.062326} {\bibfield
  {journal} {\bibinfo  {journal} {Phys. Rev. A}\ }\textbf {\bibinfo {volume}
  {100}},\ \bibinfo {pages} {062326} (\bibinfo {year} {2019})}\BibitemShut
  {NoStop}%
\bibitem [{\citenamefont {Saffman}(2016)}]{Saffman_2016}%
  \BibitemOpen
  \bibfield  {author} {\bibinfo {author} {\bibfnamefont {M.}~\bibnamefont
  {Saffman}},\ }\href {\doibase 10.1088/0953-4075/49/20/202001} {\bibfield
  {journal} {\bibinfo  {journal} {Journal of Physics B}\ }\textbf {\bibinfo
  {volume} {49}},\ \bibinfo {pages} {202001} (\bibinfo {year}
  {2016})}\BibitemShut {NoStop}%
\bibitem [{\citenamefont {Shen}\ \emph {et~al.}(2019)\citenamefont {Shen},
  \citenamefont {Wu}, \citenamefont {Su},\ and\ \citenamefont
  {Liang}}]{Shen_19}%
  \BibitemOpen
  \bibfield  {author} {\bibinfo {author} {\bibfnamefont {C.-P.}\ \bibnamefont
  {Shen}}, \bibinfo {author} {\bibfnamefont {J.-L.}\ \bibnamefont {Wu}},
  \bibinfo {author} {\bibfnamefont {S.-L.}\ \bibnamefont {Su}}, \ and\ \bibinfo
  {author} {\bibfnamefont {E.}~\bibnamefont {Liang}},\ }\href {\doibase
  10.1364/OL.44.002036} {\bibfield  {journal} {\bibinfo  {journal} {Opt.
  Lett.}\ }\textbf {\bibinfo {volume} {44}},\ \bibinfo {pages} {2036} (\bibinfo
  {year} {2019})}\BibitemShut {NoStop}%
\bibitem [{\citenamefont {Beterov}\ \emph {et~al.}(2018)\citenamefont
  {Beterov}, \citenamefont {Ashkarin}, \citenamefont {Yakshina}, \citenamefont
  {Tretyakov}, \citenamefont {Entin}, \citenamefont {Ryabtsev}, \citenamefont
  {Cheinet}, \citenamefont {Pillet},\ and\ \citenamefont
  {Saffman}}]{Beterov_2018}%
  \BibitemOpen
  \bibfield  {author} {\bibinfo {author} {\bibfnamefont {I.~I.}\ \bibnamefont
  {Beterov}}, \bibinfo {author} {\bibfnamefont {I.~N.}\ \bibnamefont
  {Ashkarin}}, \bibinfo {author} {\bibfnamefont {E.~A.}\ \bibnamefont
  {Yakshina}}, \bibinfo {author} {\bibfnamefont {D.~B.}\ \bibnamefont
  {Tretyakov}}, \bibinfo {author} {\bibfnamefont {V.~M.}\ \bibnamefont
  {Entin}}, \bibinfo {author} {\bibfnamefont {I.~I.}\ \bibnamefont {Ryabtsev}},
  \bibinfo {author} {\bibfnamefont {P.}~\bibnamefont {Cheinet}}, \bibinfo
  {author} {\bibfnamefont {P.}~\bibnamefont {Pillet}}, \ and\ \bibinfo {author}
  {\bibfnamefont {M.}~\bibnamefont {Saffman}},\ }\href {\doibase
  10.1103/PhysRevA.98.042704} {\bibfield  {journal} {\bibinfo  {journal} {Phys.
  Rev. A}\ }\textbf {\bibinfo {volume} {98}},\ \bibinfo {pages} {042704}
  (\bibinfo {year} {2018})}\BibitemShut {NoStop}%
\bibitem [{\citenamefont {Levine}\ \emph {et~al.}(2019)\citenamefont {Levine},
  \citenamefont {Keesling}, \citenamefont {Semeghini}, \citenamefont {Omran},
  \citenamefont {Wang}, \citenamefont {Ebadi}, \citenamefont {Bernien},
  \citenamefont {Greiner}, \citenamefont {Vuleti\ifmmode~\acute{c}\else
  \'{c}\fi{}}, \citenamefont {Pichler},\ and\ \citenamefont
  {Lukin}}]{Levine_2019}%
  \BibitemOpen
  \bibfield  {author} {\bibinfo {author} {\bibfnamefont {H.}~\bibnamefont
  {Levine}}, \bibinfo {author} {\bibfnamefont {A.}~\bibnamefont {Keesling}},
  \bibinfo {author} {\bibfnamefont {G.}~\bibnamefont {Semeghini}}, \bibinfo
  {author} {\bibfnamefont {A.}~\bibnamefont {Omran}}, \bibinfo {author}
  {\bibfnamefont {T.~T.}\ \bibnamefont {Wang}}, \bibinfo {author}
  {\bibfnamefont {S.}~\bibnamefont {Ebadi}}, \bibinfo {author} {\bibfnamefont
  {H.}~\bibnamefont {Bernien}}, \bibinfo {author} {\bibfnamefont
  {M.}~\bibnamefont {Greiner}}, \bibinfo {author} {\bibfnamefont
  {V.}~\bibnamefont {Vuleti\ifmmode~\acute{c}\else \'{c}\fi{}}}, \bibinfo
  {author} {\bibfnamefont {H.}~\bibnamefont {Pichler}}, \ and\ \bibinfo
  {author} {\bibfnamefont {M.~D.}\ \bibnamefont {Lukin}},\ }\href {\doibase
  10.1103/PhysRevLett.123.170503} {\bibfield  {journal} {\bibinfo  {journal}
  {Phys. Rev. Lett.}\ }\textbf {\bibinfo {volume} {123}},\ \bibinfo {pages}
  {170503} (\bibinfo {year} {2019})}\BibitemShut {NoStop}%
\bibitem [{\citenamefont {Saffman}\ and\ \citenamefont
  {M\o{}lmer}(2009)}]{Saffman_2009}%
  \BibitemOpen
  \bibfield  {author} {\bibinfo {author} {\bibfnamefont {M.}~\bibnamefont
  {Saffman}}\ and\ \bibinfo {author} {\bibfnamefont {K.}~\bibnamefont
  {M\o{}lmer}},\ }\href {\doibase 10.1103/PhysRevLett.102.240502} {\bibfield
  {journal} {\bibinfo  {journal} {Phys. Rev. Lett.}\ }\textbf {\bibinfo
  {volume} {102}},\ \bibinfo {pages} {240502} (\bibinfo {year}
  {2009})}\BibitemShut {NoStop}%
\bibitem [{\citenamefont {Su}(2018)}]{Su_2018}%
  \BibitemOpen
  \bibfield  {author} {\bibinfo {author} {\bibfnamefont {S.~L.}\ \bibnamefont
  {Su}},\ }\href {\doibase 10.1088/1674-1056/27/11/110304} {\bibfield
  {journal} {\bibinfo  {journal} {Chinese Physics B}\ }\textbf {\bibinfo
  {volume} {27}},\ \bibinfo {pages} {110304} (\bibinfo {year}
  {2018})}\BibitemShut {NoStop}%
\bibitem [{\citenamefont {{Young}}\ \emph {et~al.}(2020)\citenamefont
  {{Young}}, \citenamefont {{Bienias}}, \citenamefont {{Belyansky}},
  \citenamefont {{Kaufman}},\ and\ \citenamefont
  {{Gorshkov}}}]{2020arXiv200602486Y}%
  \BibitemOpen
  \bibfield  {author} {\bibinfo {author} {\bibfnamefont {J.~T.}\ \bibnamefont
  {{Young}}}, \bibinfo {author} {\bibfnamefont {P.}~\bibnamefont {{Bienias}}},
  \bibinfo {author} {\bibfnamefont {R.}~\bibnamefont {{Belyansky}}}, \bibinfo
  {author} {\bibfnamefont {A.~M.}\ \bibnamefont {{Kaufman}}}, \ and\ \bibinfo
  {author} {\bibfnamefont {A.~V.}\ \bibnamefont {{Gorshkov}}},\ }\href@noop {}
  {\bibfield  {journal} {\bibinfo  {journal} {arXiv e-prints}\ ,\ \bibinfo
  {eid} {arXiv:2006.02486}} (\bibinfo {year} {2020})},\ \Eprint
  {http://arxiv.org/abs/2006.02486} {arXiv:2006.02486 [quant-ph]} \BibitemShut
  {NoStop}%
\bibitem [{\citenamefont {Wu}\ \emph {et~al.}(2010)\citenamefont {Wu},
  \citenamefont {Yang},\ and\ \citenamefont {Zheng}}]{Wu_2010}%
  \BibitemOpen
  \bibfield  {author} {\bibinfo {author} {\bibfnamefont {H.-Z.}\ \bibnamefont
  {Wu}}, \bibinfo {author} {\bibfnamefont {Z.-B.}\ \bibnamefont {Yang}}, \ and\
  \bibinfo {author} {\bibfnamefont {S.-B.}\ \bibnamefont {Zheng}},\ }\href
  {\doibase 10.1103/PhysRevA.82.034307} {\bibfield  {journal} {\bibinfo
  {journal} {Phys. Rev. A}\ }\textbf {\bibinfo {volume} {82}},\ \bibinfo
  {pages} {034307} (\bibinfo {year} {2010})}\BibitemShut {NoStop}%
\bibitem [{\citenamefont {Brion}\ \emph {et~al.}(2007)\citenamefont {Brion},
  \citenamefont {Mouritzen},\ and\ \citenamefont {M\o{}lmer}}]{Brion_2007}%
  \BibitemOpen
  \bibfield  {author} {\bibinfo {author} {\bibfnamefont {E.}~\bibnamefont
  {Brion}}, \bibinfo {author} {\bibfnamefont {A.~S.}\ \bibnamefont
  {Mouritzen}}, \ and\ \bibinfo {author} {\bibfnamefont {K.}~\bibnamefont
  {M\o{}lmer}},\ }\href {\doibase 10.1103/PhysRevA.76.022334} {\bibfield
  {journal} {\bibinfo  {journal} {Phys. Rev. A}\ }\textbf {\bibinfo {volume}
  {76}},\ \bibinfo {pages} {022334} (\bibinfo {year} {2007})}\BibitemShut
  {NoStop}%
\bibitem [{\citenamefont {Carr}\ and\ \citenamefont
  {Saffman}(2013)}]{Carr_2013}%
  \BibitemOpen
  \bibfield  {author} {\bibinfo {author} {\bibfnamefont {A.~W.}\ \bibnamefont
  {Carr}}\ and\ \bibinfo {author} {\bibfnamefont {M.}~\bibnamefont {Saffman}},\
  }\href {\doibase 10.1103/PhysRevLett.111.033607} {\bibfield  {journal}
  {\bibinfo  {journal} {Phys. Rev. Lett.}\ }\textbf {\bibinfo {volume} {111}},\
  \bibinfo {pages} {033607} (\bibinfo {year} {2013})}\BibitemShut {NoStop}%
\bibitem [{\citenamefont {Zeiher}\ \emph {et~al.}(2015)\citenamefont {Zeiher},
  \citenamefont {Schau\ss{}}, \citenamefont {Hild}, \citenamefont {Macr\`{\i}},
  \citenamefont {Bloch},\ and\ \citenamefont {Gross}}]{PhysRevX.5.031015}%
  \BibitemOpen
  \bibfield  {author} {\bibinfo {author} {\bibfnamefont {J.}~\bibnamefont
  {Zeiher}}, \bibinfo {author} {\bibfnamefont {P.}~\bibnamefont {Schau\ss{}}},
  \bibinfo {author} {\bibfnamefont {S.}~\bibnamefont {Hild}}, \bibinfo {author}
  {\bibfnamefont {T.}~\bibnamefont {Macr\`{\i}}}, \bibinfo {author}
  {\bibfnamefont {I.}~\bibnamefont {Bloch}}, \ and\ \bibinfo {author}
  {\bibfnamefont {C.}~\bibnamefont {Gross}},\ }\href {\doibase
  10.1103/PhysRevX.5.031015} {\bibfield  {journal} {\bibinfo  {journal} {Phys.
  Rev. X}\ }\textbf {\bibinfo {volume} {5}},\ \bibinfo {pages} {031015}
  (\bibinfo {year} {2015})}\BibitemShut {NoStop}%
\bibitem [{\citenamefont {Weber}\ \emph {et~al.}(2015)\citenamefont {Weber},
  \citenamefont {Höning}, \citenamefont {Niederprüm}, \citenamefont
  {Manthey}, \citenamefont {Thomas}, \citenamefont {Guarrera}, \citenamefont
  {Fleischhauer}, \citenamefont {Barontini},\ and\ \citenamefont
  {Ott}}]{np.11.157}%
  \BibitemOpen
  \bibfield  {author} {\bibinfo {author} {\bibfnamefont {T.~M.}\ \bibnamefont
  {Weber}}, \bibinfo {author} {\bibfnamefont {M.}~\bibnamefont {Höning}},
  \bibinfo {author} {\bibfnamefont {T.}~\bibnamefont {Niederprüm}}, \bibinfo
  {author} {\bibfnamefont {T.}~\bibnamefont {Manthey}}, \bibinfo {author}
  {\bibfnamefont {O.}~\bibnamefont {Thomas}}, \bibinfo {author} {\bibfnamefont
  {V.}~\bibnamefont {Guarrera}}, \bibinfo {author} {\bibfnamefont
  {M.}~\bibnamefont {Fleischhauer}}, \bibinfo {author} {\bibfnamefont
  {G.}~\bibnamefont {Barontini}}, \ and\ \bibinfo {author} {\bibfnamefont
  {H.}~\bibnamefont {Ott}},\ }\href@noop {} {\bibfield  {journal} {\bibinfo
  {journal} {Nature Physics}\ }\textbf {\bibinfo {volume} {11}},\ \bibinfo
  {pages} {157} (\bibinfo {year} {2015})}\BibitemShut {NoStop}%
\bibitem [{\citenamefont {Shi}(2018)}]{Shi_2018}%
  \BibitemOpen
  \bibfield  {author} {\bibinfo {author} {\bibfnamefont {X.-F.}\ \bibnamefont
  {Shi}},\ }\href {\doibase 10.1103/PhysRevApplied.9.051001} {\bibfield
  {journal} {\bibinfo  {journal} {Phys. Rev. Applied}\ }\textbf {\bibinfo
  {volume} {9}},\ \bibinfo {pages} {051001} (\bibinfo {year}
  {2018})}\BibitemShut {NoStop}%
\bibitem [{\citenamefont {Isenhower}\ \emph {et~al.}(2011)\citenamefont
  {Isenhower}, \citenamefont {Saffman},\ and\ \citenamefont
  {M\o{}Lmer}}]{Isenhower_2011}%
  \BibitemOpen
  \bibfield  {author} {\bibinfo {author} {\bibfnamefont {L.}~\bibnamefont
  {Isenhower}}, \bibinfo {author} {\bibfnamefont {M.}~\bibnamefont {Saffman}},
  \ and\ \bibinfo {author} {\bibfnamefont {K.}~\bibnamefont {M\o{}Lmer}},\
  }\href {\doibase doi.org/10.1007/s11128-011-0292-4} {\bibfield  {journal}
  {\bibinfo  {journal} {Quantum Information Processing}\ }\textbf {\bibinfo
  {volume} {10}},\ \bibinfo {pages} {755} (\bibinfo {year} {2011})}\BibitemShut
  {NoStop}%
\bibitem [{\citenamefont {Petrosyan}\ and\ \citenamefont
  {M\o{}lmer}(2014)}]{Petrosyan_2014}%
  \BibitemOpen
  \bibfield  {author} {\bibinfo {author} {\bibfnamefont {D.}~\bibnamefont
  {Petrosyan}}\ and\ \bibinfo {author} {\bibfnamefont {K.}~\bibnamefont
  {M\o{}lmer}},\ }\href {\doibase 10.1103/PhysRevLett.113.123003} {\bibfield
  {journal} {\bibinfo  {journal} {Phys. Rev. Lett.}\ }\textbf {\bibinfo
  {volume} {113}},\ \bibinfo {pages} {123003} (\bibinfo {year}
  {2014})}\BibitemShut {NoStop}%
\bibitem [{\citenamefont {M{\o}lmer}\ \emph {et~al.}(1993)\citenamefont
  {M{\o}lmer}, \citenamefont {Castin},\ and\ \citenamefont
  {Dalibard}}]{Molmer_1993}%
  \BibitemOpen
  \bibfield  {author} {\bibinfo {author} {\bibfnamefont {K.}~\bibnamefont
  {M{\o}lmer}}, \bibinfo {author} {\bibfnamefont {Y.}~\bibnamefont {Castin}}, \
  and\ \bibinfo {author} {\bibfnamefont {J.}~\bibnamefont {Dalibard}},\ }\href
  {\doibase 10.1364/JOSAB.10.000524} {\bibfield  {journal} {\bibinfo  {journal}
  {J. Opt. Soc. Am. B}\ }\textbf {\bibinfo {volume} {10}},\ \bibinfo {pages}
  {524} (\bibinfo {year} {1993})}\BibitemShut {NoStop}%
\bibitem [{\citenamefont {Beterov}\ \emph {et~al.}(2009)\citenamefont
  {Beterov}, \citenamefont {Ryabtsev}, \citenamefont {Tretyakov},\ and\
  \citenamefont {Entin}}]{Beterov_2009}%
  \BibitemOpen
  \bibfield  {author} {\bibinfo {author} {\bibfnamefont {I.~I.}\ \bibnamefont
  {Beterov}}, \bibinfo {author} {\bibfnamefont {I.~I.}\ \bibnamefont
  {Ryabtsev}}, \bibinfo {author} {\bibfnamefont {D.~B.}\ \bibnamefont
  {Tretyakov}}, \ and\ \bibinfo {author} {\bibfnamefont {V.~M.}\ \bibnamefont
  {Entin}},\ }\href {\doibase 10.1103/PhysRevA.79.052504} {\bibfield  {journal}
  {\bibinfo  {journal} {Phys. Rev. A}\ }\textbf {\bibinfo {volume} {79}},\
  \bibinfo {pages} {052504} (\bibinfo {year} {2009})}\BibitemShut {NoStop}%
\bibitem [{\citenamefont {Zhang}\ \emph {et~al.}(2012)\citenamefont {Zhang},
  \citenamefont {Gill}, \citenamefont {Isenhower}, \citenamefont {Walker},\
  and\ \citenamefont {Saffman}}]{PhysRevA.85.042310}%
  \BibitemOpen
  \bibfield  {author} {\bibinfo {author} {\bibfnamefont {X.~L.}\ \bibnamefont
  {Zhang}}, \bibinfo {author} {\bibfnamefont {A.~T.}\ \bibnamefont {Gill}},
  \bibinfo {author} {\bibfnamefont {L.}~\bibnamefont {Isenhower}}, \bibinfo
  {author} {\bibfnamefont {T.~G.}\ \bibnamefont {Walker}}, \ and\ \bibinfo
  {author} {\bibfnamefont {M.}~\bibnamefont {Saffman}},\ }\href {\doibase
  10.1103/PhysRevA.85.042310} {\bibfield  {journal} {\bibinfo  {journal} {Phys.
  Rev. A}\ }\textbf {\bibinfo {volume} {85}},\ \bibinfo {pages} {042310}
  (\bibinfo {year} {2012})}\BibitemShut {NoStop}%
\bibitem [{\citenamefont {Graham}\ \emph {et~al.}(2019)\citenamefont {Graham},
  \citenamefont {Kwon}, \citenamefont {Grinkemeyer}, \citenamefont {Marra},
  \citenamefont {Jiang}, \citenamefont {Lichtman}, \citenamefont {Sun},
  \citenamefont {Ebert},\ and\ \citenamefont {Saffman}}]{Graham_2019}%
  \BibitemOpen
  \bibfield  {author} {\bibinfo {author} {\bibfnamefont {T.~M.}\ \bibnamefont
  {Graham}}, \bibinfo {author} {\bibfnamefont {M.}~\bibnamefont {Kwon}},
  \bibinfo {author} {\bibfnamefont {B.}~\bibnamefont {Grinkemeyer}}, \bibinfo
  {author} {\bibfnamefont {Z.}~\bibnamefont {Marra}}, \bibinfo {author}
  {\bibfnamefont {X.}~\bibnamefont {Jiang}}, \bibinfo {author} {\bibfnamefont
  {M.~T.}\ \bibnamefont {Lichtman}}, \bibinfo {author} {\bibfnamefont
  {Y.}~\bibnamefont {Sun}}, \bibinfo {author} {\bibfnamefont {M.}~\bibnamefont
  {Ebert}}, \ and\ \bibinfo {author} {\bibfnamefont {M.}~\bibnamefont
  {Saffman}},\ }\href {\doibase 10.1103/PhysRevLett.123.230501} {\bibfield
  {journal} {\bibinfo  {journal} {Phys. Rev. Lett.}\ }\textbf {\bibinfo
  {volume} {123}},\ \bibinfo {pages} {230501} (\bibinfo {year}
  {2019})}\BibitemShut {NoStop}%
\bibitem [{\citenamefont {M\"uller}\ \emph {et~al.}(2009)\citenamefont
  {M\"uller}, \citenamefont {Lesanovsky}, \citenamefont {Weimer}, \citenamefont
  {B\"uchler},\ and\ \citenamefont {Zoller}}]{PhysRevLett.102.170502}%
  \BibitemOpen
  \bibfield  {author} {\bibinfo {author} {\bibfnamefont {M.}~\bibnamefont
  {M\"uller}}, \bibinfo {author} {\bibfnamefont {I.}~\bibnamefont
  {Lesanovsky}}, \bibinfo {author} {\bibfnamefont {H.}~\bibnamefont {Weimer}},
  \bibinfo {author} {\bibfnamefont {H.~P.}\ \bibnamefont {B\"uchler}}, \ and\
  \bibinfo {author} {\bibfnamefont {P.}~\bibnamefont {Zoller}},\ }\href
  {\doibase 10.1103/PhysRevLett.102.170502} {\bibfield  {journal} {\bibinfo
  {journal} {Phys. Rev. Lett.}\ }\textbf {\bibinfo {volume} {102}},\ \bibinfo
  {pages} {170502} (\bibinfo {year} {2009})}\BibitemShut {NoStop}%
\bibitem [{\citenamefont {Jaksch}\ \emph {et~al.}(2000)\citenamefont {Jaksch},
  \citenamefont {Cirac}, \citenamefont {Zoller}, \citenamefont {Rolston},
  \citenamefont {C\^ot\'e},\ and\ \citenamefont {Lukin}}]{PhysRevLett.85.2208}%
  \BibitemOpen
  \bibfield  {author} {\bibinfo {author} {\bibfnamefont {D.}~\bibnamefont
  {Jaksch}}, \bibinfo {author} {\bibfnamefont {J.~I.}\ \bibnamefont {Cirac}},
  \bibinfo {author} {\bibfnamefont {P.}~\bibnamefont {Zoller}}, \bibinfo
  {author} {\bibfnamefont {S.~L.}\ \bibnamefont {Rolston}}, \bibinfo {author}
  {\bibfnamefont {R.}~\bibnamefont {C\^ot\'e}}, \ and\ \bibinfo {author}
  {\bibfnamefont {M.~D.}\ \bibnamefont {Lukin}},\ }\href {\doibase
  10.1103/PhysRevLett.85.2208} {\bibfield  {journal} {\bibinfo  {journal}
  {Phys. Rev. Lett.}\ }\textbf {\bibinfo {volume} {85}},\ \bibinfo {pages}
  {2208} (\bibinfo {year} {2000})}\BibitemShut {NoStop}%
\end{thebibliography}%


\end{document}